\documentclass[aps,prl,twocolumn,groupedaddress,nofootinbib,showpacs]{revtex4}
\usepackage[dvips]{graphicx}
\newcommand{\BlackHat}{{\sc BlackHat}}
\newcommand{\SHERPA}{{\sc SHERPA}}
\newcommand{\MCFM}{{\sc MCFM}}
\newcommand{\AMEGIC}{{\sc AMEGIC++}}

\newif\ifdraft
\drafttrue
\newif\ifpreprint
\preprintfalse

\def\fig#1{fig.~{\ref{#1}}}
\def\Fig#1{Fig.~{\ref{#1}}}

\def\tab#1{table~{\ref{#1}}}

\def\Wjjj{$W\,\!+\,3$}
\def\Wjjjj{$W\,\!+\,4$}

\def\WZjj{$W,Z\,\!+\,2$}

\def\Zz{$Z\,\!+\,0$}
\def\Zj{$Z\,\!+\,1$}

\def\Zjjjj{$Z\,\!+\,4$}
\def\Zjn{$Z\,\!+\,n$}
\def\Zjnm{$Z\,\!+\,(n\!-\!1)$}
\def\Zgamjn{$Z,\gamma^*\,\!+\,n$}

\def\Zgamjjjj{$Z,\gamma^*\,\!+\,4$}
\def\Zjjjj{$Z\,\!+\,4$}

\def\WZjjjj{$W,Z/\gamma^*\,\!+\,4$}
\def\jet{{\rm jet}}
\def\pt{p_T}

\def\ET{E_T}

\def\HTpartonicp{{\hat H}_T'}

\newbox\charbox
\newbox\slabox
\def\s#1{{      
        \setbox\charbox=\hbox{$#1$}
        \setbox\slabox=\hbox{$/$}
        \dimen\charbox=\ht\slabox
        \advance\dimen\charbox by -\dp\slabox
        \advance\dimen\charbox by -\ht\charbox
        \advance\dimen\charbox by \dp\charbox
        \divide\dimen\charbox by 2
        \raise-\dimen\charbox\hbox to \wd\charbox{\hss/\hss}
        \llap{$#1$}
}}

\begin{document}

\title{
\ifpreprint
\hbox{\rm\small
SB/F/390-11$\null\hskip 2.2cm \null$
UCLA/11/TEP/109$\null\hskip 2.2cm \null$
SLAC--PUB--14527$\null\hskip 2.4cm \null$
NSF-KITP-11-167

\break}
\hbox{\rm\small IPPP/11/45 $\null\hskip 4.4cm \null$
 $\null\hskip 7cm \null$
 CERN--PH--TH/2011/188\break}
\hbox{$\null$\break}
\fi
Precise Predictions for $Z + 4$ Jets at Hadron Colliders
}

\author{H.~Ita${}^a$,\, Z.~Bern${}^a$,\,
L.~J.~Dixon${}^{b,c}$,\, F.~Febres Cordero${}^d$
,\,D.~A.~Kosower${}^{e}$\, and D.~Ma\^{\i}tre${}^{b,f}$ 
\\
$\null$
\\
${}^a$Department of Physics and Astronomy, UCLA, Los Angeles, CA
90095-1547, USA \\
${}^b$Theory Division, Physics Department, CERN, CH--1211 Geneva 23, 
    Switzerland\\
${}^c$SLAC National Accelerator Laboratory, Stanford University,
             Stanford, CA 94309, USA \\
${}^d$Departamento de F\'{\i}sica, Universidad Sim\'on Bol\'{\i}var, 
 Caracas 1080A, Venezuela\\
${}^e$Institut de Physique Th\'eorique, CEA--Saclay,
          F--91191 Gif-sur-Yvette cedex, France\\
${}^f$Department of Physics, University of Durham,
          Durham DH1 3LE, UK\\
}

\begin{abstract}
We present the cross section for production of a $Z$ boson in
association with four jets at the Large Hadron Collider,
at next-to-leading order in the QCD coupling.
When the $Z$ decays to neutrinos, this process is a key
irreducible background to many searches for new physics.
Its computation has been made feasible through the development
of the on-shell approach to perturbative quantum field theory.
We present the total cross section for $pp$ collisions at $\sqrt{s}=7$~TeV, 
after folding in the decay of the $Z$ boson, or virtual photon, to
a charged-lepton pair.  We also provide distributions of the transverse
momenta of the four jets, and we compare cross sections and distributions to
the corresponding ones for the production of a $W$ boson with
accompanying jets.
\end{abstract}

\pacs{12.38.-t, 12.38.Bx, 13.87.-a, 14.70.Hp \hspace{1cm}}

\maketitle


The Large Hadron Collider (LHC) is currently extending the energy
frontier into uncharted territory, in the quest to identify new
physics beyond the Standard Model of particle physics.  Many
signals of new physics, especially those containing dark matter candidates,
lie in broad distributions with significant Standard Model backgrounds.
A first-principles understanding of these backgrounds is provided by
quantum chromodynamics (QCD) and the QCD-improved parton model.
The leading perturbative order (LO) in the QCD
coupling $\alpha_s$ gives a good qualitative prediction.
Quantitatively reliable predictions require, at the least, 
next-to-leading-order (NLO) accuracy in the QCD coupling. For processes
at a hadron collider with many-jet final states, NLO computations have
long been a formidable challenge to particle theorists.

\begin{figure}[t]
\includegraphics[clip,scale=0.45]{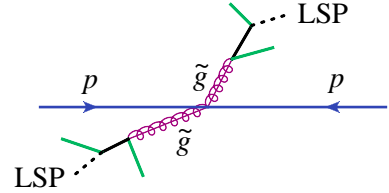}
\caption{Gluino pair production illustrates a typical signature of 
new physics scenarios: four jets plus a pair of lightest supersymmetric
particles (LSPs) that escape the detector, yielding missing
transverse energy.}
\label{GluinoFigure}
\end{figure}

In this article we present the first NLO QCD results for $Z$ boson
production in association with four jets at a hadron collider,
specifically at the LHC.  
We fold in the decay of the $Z$ boson to
an $e^+e^-$ pair (or equivalently $\mu^+\mu^-$), and include
contributions from virtual-photon exchange (collectively
denoted by $Z,\gamma^*$).  This process, containing
identifiable charged leptons, is a benchmark for the closely related
process in which the $Z$ decays into neutrinos, which appear as
missing transverse energy.  The $Z\to\nu\bar\nu$ decay mode generates a key
background process in the search for supersymmetry, as well as for
other models that lead to dark-matter particle production at the end
of a cascade of strongly-produced new particles.  \Fig{GluinoFigure}
shows a typical signal process, leading to the same signature of missing
transverse energy with four jets and no sharp resonance.  We note
that another approach to estimating this process --- combining a
measurement of prompt-photon production with a theoretical estimate of
the $Z$-to-photon
ratio~\cite{CMSZPhoton,ATLASZPhoton,CambridgeZPhoton} --- also
benefits from NLO cross sections~\cite{PhotonZ}.

\begin{figure}[t]
\includegraphics[clip,scale=0.37]{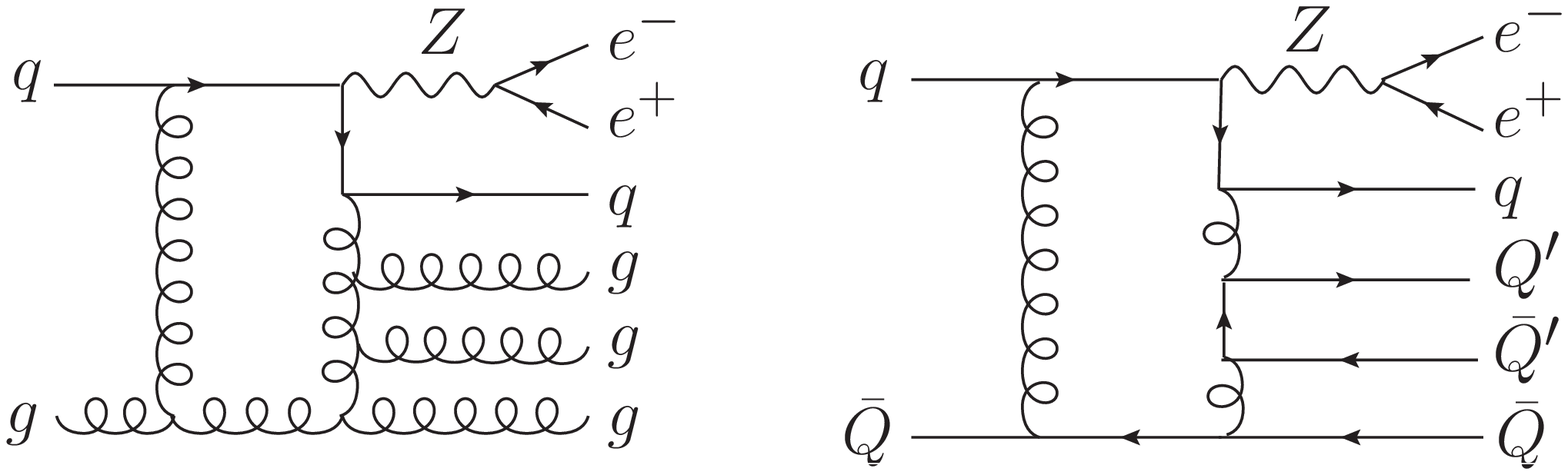}
\caption{Sample diagrams for the seven-point loop amplitudes for $q g
  \rightarrow Z q g g g$ and $q \bar Q \rightarrow Z q Q' \bar Q'\bar Q$,
  followed by $Z \to e^+ e^-$.  There are also small contributions
  where the $Z$ boson is replaced by a photon. This process is
  very similar theoretically to the case $Z \to \nu \bar \nu$
  with missing transverse energy. }
\label{lcdiagramsFigure}
\end{figure}

\begin{table*}
\vskip .4 cm
\begin{tabular}{||c||c|c||c|c||c|c||}
\hline
no. jets &  $Z$ LO  & $Z$ NLO & $Z/W^+$ LO &  $Z/W^+$ NLO & $Z n/(n\!-\!1)$ LO & $Z n/(n\!-\!1)$ NLO  \\
\hline
0 & $323.1(0.1)^{+39.3}_{-44.3}$  & $428.6(0.3)^{+6.2}_{-4.1}$ &
    $0.1209(0.0001)$  &  $0.1306(0.0003) $ & ---  & ---  
 \\
\hline
1 & $\;66.69(0.04)^{+5.59}_{-5.30} \;$ &   $ \;82.1(0.1)^{+3.3}_{-2.6} \; $ &
     $\;0.1674(0.0002)\;$ & $\; 0.166(0.001)\;$ & 
     $\;0.2064(0.0001)\; $ &
     $\; 0.1915(0.0004)\; $ 
 \\
\hline
2 & $\;19.10(0.02)^{+5.32}_{-3.82} \;$ & $20.25(0.07)^{+0.31}_{-1.02}$ & 
      $0.1636(0.0003)$  & $\;0.166(0.002)\;$ &
      $\;0.2864(0.0003)\; $& 
      $0.247(0.001)$
 \\
\hline
3 & $\;4.76(0.01)^{+2.18}_{-1.35} \;$  &  $\;4.73(0.03)^{+0.05}_{-0.35}\;$ &
    $\; 0.1634(0.0004) \; $  &  $\;0.169(0.002)\;$ & 
    $\; 0.2494(0.0004)\; $ & 
    $0.234(0.002)$
 \\
\hline
4 & $\;1.116(0.002)^{+0.695}_{-0.390} \;$  &  $ \;1.06(0.01)^{+0.05}_{-0.14}\; $ &
    $0.1618(0.0003)$  & $0.172(0.002)$ &
    $0.2343(0.0005) $ & 
    $0.223(0.002)$
 \\
\hline 
\end{tabular} 
\caption{Total cross sections in pb for \Zgamjn-jet production at
  the LHC, using the anti-$k_T$ jet algorithm with
  $R=0.5$.  The NLO result for \Zgamjjjj{} jets uses the leading-color
  virtual approximation.  The fourth and fifth columns
  give the cross-section ratios for $Z,\gamma^*\rightarrow e^+ e^-$
  production to $W^+ \rightarrow e^+\nu$ production.  The final two
  columns give the ratios of the cross section for the given process
  to that with one fewer jet.  The numerical integration uncertainty is
  in parentheses.  The scale dependence is given in superscripts and
  subscripts. 
\label{CrossSectionAnti-kt-R5Table} 
}
\end{table*}

Recent years have witnessed a growing number of NLO QCD results using
both traditional and on-shell
approaches~\cite{PRLW3BH,EMZW3Tev,W3jDistributions,
MZ3j,TeVZ,OtherNLO,W4j}.  On-shell
methods~\cite{UnitarityMethod,BCFW,Bootstrap,OPPEtc} exploit the
analytic properties that all scattering amplitudes must satisfy, and
generate new amplitudes from previously-computed ones.
Computationally, they scale modestly with increasing numbers of
external partons.  We used these methods to compute the production of
a $W$ or $Z$ boson in association with three jets at
NLO~\cite{W3jDistributions,TeVZ}.  The predictions are generally in
very good agreement with data from the
Tevatron~\cite{D0W123,TevatronPapers}.  (Earlier NLO results for
\Wjjj{} jets were based on similar techniques and used various
leading-color approximations~\cite{PRLW3BH,EMZW3Tev,MZ3j}.)  We have
also calculated~\cite{W4j} \Wjjjj-jet production at the LHC, making
use of a leading-color approximation for the virtual terms that is
known to be valid to about 3\% for up to three associated
jets~\cite{W3jDistributions,TeVZ}.  We will use the same approximation
for \Zjjjj{}-jet production.  The \Zjjjj{}-jet production computation
is significantly more complex than that for \Wjjjj{}-jet production,
because the quark flavor structure leads to more partonic
subprocesses, especially those containing identical fermion pairs.

\begin{figure*}[t]
\includegraphics[clip,scale=0.52]{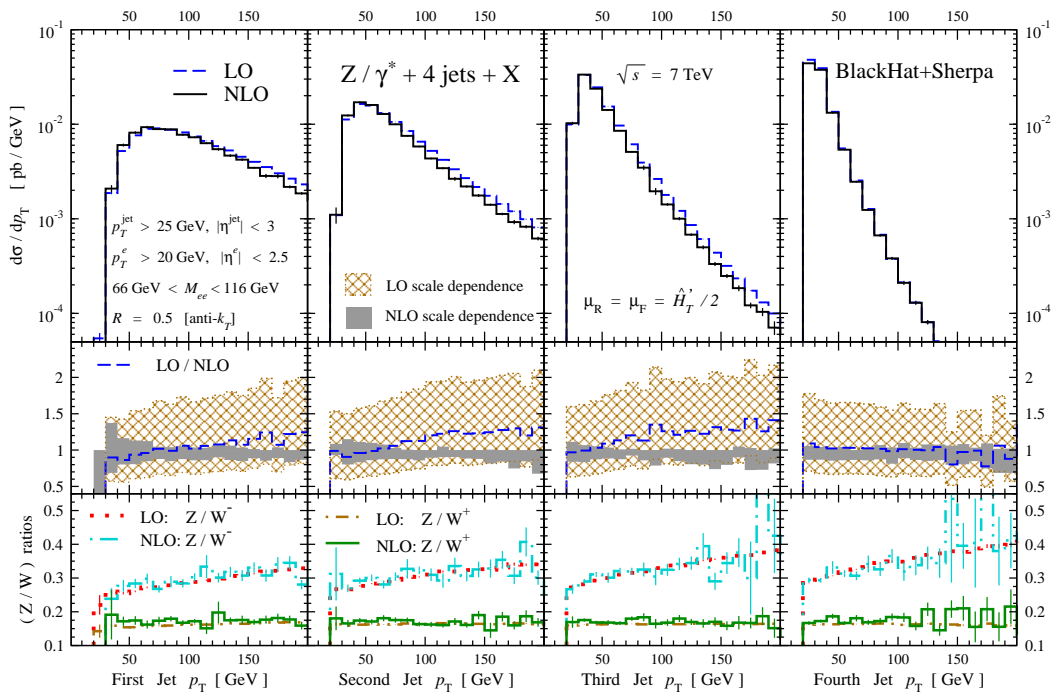}
\caption{A comparison of the $p_T$ distributions of the leading four
  jets in \Zgamjjjj{}-jet production at the LHC.
  In the upper panels the NLO distribution is the solid (black)
  histogram and the LO predictions are shown as dashed (blue) lines.
  The thin vertical line in the center of each bin (where visible)
  gives its numerical (Monte Carlo) integration error.  The middle
  panels show the LO distribution and LO and NLO scale-dependence
  bands normalized to the central NLO prediction.  The bands are
  shaded (gray) for NLO and cross-hatched (brown) for LO.  In the
  bottom panel, the dotted (red) line is the LO $Z/W^-$ ratio, the
  dot-longer-dash (cyan) line the NLO $Z/W^-$ ratio, the
  dot-shorter-dash (brown) line the LO $Z/W^+$ ratio and the 
  solid (green) line the NLO $Z/W^+$ ratio. }
\label{Z4ptFigure}
\end{figure*}

We use the same basic setup as in our earlier
work~\cite{W3jDistributions,TeVZ}.  Virtual
contributions are evaluated via the \BlackHat{} package~\cite{BlackHatI},
an implementation of on-shell methods.
We incorporate a number of significant improvements in automating the
assembly of subprocesses
and in ensuring the numerical stability of the virtual corrections.
To minimize the amount of higher-precision recomputation at points
for which an instability is detected, only the unstable part, rather
than the whole matrix element, is
recomputed~\cite{W3jDistributions,HaraldReview}.
Representative virtual diagrams are shown in \fig{lcdiagramsFigure}.
We include all subprocesses, and
make the leading-color virtual approximation only in \WZjjjj-jet
production.  As in ref.~\cite{TeVZ}, we drop small axial and vector
loop contributions, along with the small effects of
top quarks in the loop.

The remaining NLO ingredients, the real-emission and
dipole-subtraction terms~\cite{CS}, are computed using
\AMEGIC{}~\cite{Amegic}, which is part of the \SHERPA{}
package~\cite{Sherpa}.  Here we retain the full color dependence.
The \SHERPA-based phase-space integration exploits QCD
antenna structures~\cite{AntennaIntegrator,GleisbergIntegrator}.
\BlackHat{} supplies the real-emission tree amplitudes, using on-shell
recursion relations~\cite{BCFW} and efficient analytic forms
extracted from ${\cal N}=4$ super-Yang-Mills
theory~\cite{DrummondTrees}.
We have validated the code extensively.  Previously, we compared many
results against \MCFM{}~\cite{MCFM} for \WZjj{}-jet production. 

Cross sections and distributions at LO suffer from strong sensitivity
to the unphysical renormalization scale $\mu_R$ and factorization scale
$\mu_F$ entering $\alpha_s$ and the parton
distributions.  This dependence is reduced at NLO.  This issue is
especially important at the LHC because of the wide range of
kinematics probed.  This wide range also obliges us to
choose an event-by-event scale characteristic of the kinematics when we
compute distributions.  We choose
$\mu_R = \mu_F \equiv \mu = \HTpartonicp/2$ as our
central scale~\cite{W4j}, where
$\HTpartonicp \equiv \sum_i p_T^i + \ET^Z$.
The sum runs over all final-state partons $i$, and 
$\ET^Z \equiv \sqrt{M_Z^2+(p_T^{e^+e^-})^2}$;
$M_Z$ is fixed to its on-shell value.
We follow standard procedure to assess scale dependence,
varying the central scale up and down by a factor of two to construct
scale-dependence bands, taking the minimum and maximum of any
observable evaluated at five values: $\mu\times (1/2, 1/\sqrt2, 1,
\sqrt2, 2)$.

The fixed-order perturbative expansion
may break down in special kinematic regions, where large
logarithms of ratios of physical scales emerge.  Threshold
logarithms can affect production at very large partonic
center-of-mass energies.  However, in ref.~\cite{W4j} it was argued,
using results for inclusive single-jet production~\cite{deFV},
that at the mass scales probed in \WZjjjj-jet production, such
logarithms should remain quite modest.
Tighter cuts can isolate regions subject to potentially large
logarithms of either QCD or electroweak origin. In particular, cuts that
force the vector boson to large $\pt$, the desired region for many searches
for supersymmetry or dark-matter particles, can induce large electroweak
Sudakov logarithms.

In our study, we consider the inclusive process $p p \rightarrow$
\Zjjjj{} jets at an LHC center-of-mass energy of $\sqrt{s} = 7$ TeV.
We incorporate the full $Z,\gamma^*$ Breit-Wigner resonance and decay
the intermediate boson into an electron--positron pair at the
amplitude level, retaining all spin correlations.  We impose the
following cuts on the transverse momenta $p_T$, and pseudorapidities
$\eta$: $\pt^{e} > 20$ GeV, $|\eta^e| < 2.5$, $\pt^\jet > 25$ GeV,
$|\eta^\jet|<3$, and 66 GeV${} < M_{e^+ e^-} < 116$ GeV.  The lower
cut on the lepton-pair invariant mass $M_{e^+ e^-}$ eliminates the
large contribution from the photon pole.  Jets are defined using the
infrared-safe anti-$k_T$ algorithm~\cite{antikT} adopted by the LHC
experiments.  Here we present results for size parameter $R = 0.5$.
We order the jets in $\pt$.  In comparisons to $W$-boson cross
sections we follow exactly the cuts of ref.~\cite{W4j}; the jet cuts
are identical.  We use the CTEQ6M~\cite{CTEQ6M} parton distribution
functions at NLO, and the CTEQ6L1 set at LO.  Electroweak boson masses
and couplings are chosen as in refs.~\cite{W3jDistributions,TeVZ}.  We
also use the SHERPA six-flavor implementation of $\alpha_s(\mu)$ and
the value of $\alpha_s(M_Z)$ provided by CTEQ.

In \tab{CrossSectionAnti-kt-R5Table}, we give LO and NLO parton-level
inclusive cross sections for $e^+e^-$ production via a $Z,\gamma^*$
boson, and accompanied by zero through four jets.  The NLO results
exhibit a markedly reduced scale dependence compared to LO; the
improvement becomes stronger as the number of jets increases.  We also
display the ratios of the $Z$ to $W^+$ cross sections, and the
``jet-production'' ratios of \Zjn-jet to \Zjnm-jet cross sections.
Ratios to $W^-$-boson cross sections can be obtained using the
results of ref.~\cite{W4j}.  Both kinds of ratios should be less
sensitive to theoretical systematics than the absolute cross sections.
Indeed, the $Z$/$W$ ratios show relatively little difference between
LO and NLO.  This ratio changes very little under correlated
variations of $\mu$ in numerator and denominator; hence we do not
exhibit such scale variation.  Varying the $R$ parameter in
the jet algorithm, we find very similar behavior as in the
$W$ case~\cite{W4j}.

It has generally been expected that the jet-production ratio is
roughly independent of the number of jets~\cite{BerendsRatio}.  Other
than the \Zj-jet/\Zz-jet ratio, which is smaller because of the
restricted kinematics of the leading contribution to \Zz-jet
production, the results shown in \tab{CrossSectionAnti-kt-R5Table} are
consistent with this expectation. 
The ratios are, however, rather sensitive to the
experimental cuts: for example, imposing large vector-boson $\pt$ cuts
makes them depend strongly on the number of jets~\cite{TeVZ}.

In \fig{Z4ptFigure}, we show the $\pt$ distributions of the leading
four jets in \Zgamjjjj-jet production at LO and NLO.
The predictions are normalized
to the central NLO prediction in the middle panels.  The NLO distributions
display a much smaller dependence on the unphysical
renormalization and factorization scales.  For our central scale
choice, the distributions for the first three leading jets soften
noticeably from LO to NLO, while the fourth-jet distribution is virtually
unchanged.  The NLO corrections to the behavior
of \Zgamjjjj-jet and \Wjjjj-jet production are quite similar in this
respect~\cite{W4j}.

The bottom panels in \fig{Z4ptFigure} show the ratio of $Z/W^+$ and
$Z/W^-$ production both at LO and at NLO. The $Z/W^-$ ratio
rises with rising $\pt$ while the $Z/W^+$ ratio is roughly flat. Both
ratios reflect the rising dominance of the $u$ quark distribution over
the $d$ quark with increasing parton fraction $x$.  Because the $Z$ has
an appreciable coupling to an initial $u$ quark (unlike the $W^-$),
the shape of the $p_T$ distribution follows more closely the $W^+$
case than the $W^-$ case, which has a $d(x)/u(x)$ relative suppression.
The excellent agreement between LO and NLO ratios for
$Z/W^\pm$ production shows that these ratios are under solid
perturbative control. 


A comparison of parton-level results to experimental data requires
estimating the size of non-perturbative effects, such as those induced
by the underlying event or by fragmentation and hadronization of the
outgoing partons.  Standard LO parton-shower Monte Carlo programs can
provide these estimates.  As NLO parton-shower programs are
developed~\cite{POWHEGBOXMCNLO}, they can use virtual corrections computed
with \BlackHat{}.  We expect non-perturbative effects to largely cancel in
the  $Z/W^\pm$ ratios.

In the present study of the \Zgamjjjj-jet process, we have imposed
cuts typical of Standard-Model measurements at the LHC.  The same code
can be used to study the size of QCD corrections for observables
under cuts used in new-physics searches.  This will allow the study of
backgrounds to missing energy signals of new physics, arising when a
$Z$ boson decays to a pair of neutrinos.  Ratios such as the $Z/W\,+\,$jets
ratios offer highly-reliable theoretical predictions. 
Applying \BlackHat{} along with \SHERPA{} brings an unprecedented
level of theoretical precision to Standard-Model backgrounds, aiding
in the hunt for new-physics signals at the LHC.


\vskip .3 cm 

We thank Giovanni Diana, Stefan H{\"o}che and Kemal Ozeren for many
helpful discussions.  We also thank Carola Berger, Darren Forde, and
Tanju Gleisberg for contributing to earlier versions of \BlackHat.  We
thank the Kavli Institute for Theoretical Physics, where part of this
work was performed, for its hospitality.  This research was supported by
the US Department of Energy under contracts DE--FG03--91ER40662,
DE--AC02--76SF00515 and DE--FC02--94ER40818.  DAK's research is
supported by the European Research Council under Advanced Investigator
Grant ERC--AD--228301.  H.I.'s work is supported by a grant from the US
LHC Theory Initiative through NSF contract PHY--0705682. This research
was also supported in part by the National Science Foundation under Grant
No. NSF PHY05-51164.  This research used resources of Academic
Technology Services at UCLA.


\begin{thebibliography}{99}

\bibitem{CMSZPhoton}
S.~Chatrchyan {\it et al.}  [CMS Collaboration],
1106.4503 [hep-ex].

\bibitem{ATLASZPhoton}
G.~Aad {\it et al.} [ATLAS Collaboration], ATLAS-CONF-2011-086.

\bibitem{CambridgeZPhoton}
S.~Ask {\it et al.}, 
1107.2803 [hep-ph].

\bibitem{PhotonZ}
Z.~Bern {\it et al.},
1106.1423 [hep-ph].

\bibitem{PRLW3BH}
C.~F.~Berger {\it et al.},
Phys.\ Rev.\ Lett.\ {\bf 102}, 222001 (2009)\ifpreprint{} [0902.2760 [hep-ph]]\fi.

\bibitem{EMZW3Tev}
R.~K.~Ellis, K.~Melnikov and G.~Zanderighi,
Phys.\ Rev.\  D {\bf 80}, 094002 (2009)\ifpreprint{} [0906.1445 [hep-ph]]\fi.

\bibitem{W3jDistributions}
C.~F.~Berger {\it et al.},
Phys.\ Rev.\  D {\bf 80}, 074036 (2009)\ifpreprint{} [0907.1984 [hep-ph]]\fi.

\bibitem{MZ3j}
K.~Melnikov and G.~Zanderighi,
Phys.\ Rev.\  D {\bf 81}, 074025 (2010)\ifpreprint{} [0910.3671 [hep-ph]]\fi.

\bibitem{TeVZ}
C.~F.~Berger {\it et al.},
Phys.\ Rev.\  D {\bf 82}, 074002 (2010)\ifpreprint{} [1004.1659 [hep-ph]]\fi.

\bibitem{OtherNLO}
A.~Bredenstein, A.~Denner, S.~Dittmaier and S.~Pozzorini,
JHEP {\bf 0808}, 108 (2008)\ifpreprint{} [0807.1248 [hep-ph]]\fi;
%
Phys.\ Rev.\ Lett.\ {\bf 103}, 012002 (2009)\ifpreprint{} [0905.0110 [hep-ph]]\fi;
%
JHEP {\bf 1003}, 021 (2010)\ifpreprint{} [1001.4006 [hep-ph]]\fi;
%
G.~Bevilacqua {\it et al.},
JHEP {\bf 0909}, 109 (2009)\ifpreprint{} [0907.4723 [hep-ph]]\fi;
%
T.~Binoth {\it et al.},
Phys.\ Lett.\  B {\bf 685}, 293 (2010)\ifpreprint{} [0910.4379 [hep-ph]]\fi;
%
G.~Bevilacqua, M.~Czakon, C.~G.~Papadopoulos and M.~Worek,
Phys.\ Rev.\ Lett.\ {\bf 104}, 162002 (2010)\ifpreprint{} [1002.4009 [hep-ph]]\fi;
T.~Melia, K.~Melnikov, R.~Rontsch and G.~Zanderighi,
Phys.\ Rev.\  D {\bf 83}, 114043 (2011)\ifpreprint{} [1104.2327 [hep-ph]]\fi;
%
F.~Campanario, C.~Englert, M.~Rauch and D.~Zeppenfeld,
1106.4009 [hep-ph].

\bibitem{W4j}
C.~F.~Berger {\it et al.},
Phys.\ Rev.\ Lett.\ {\bf 106},\hskip -3pt 092001 (2011)
\ifpreprint{} [1009.2338 [hep-ph]]\fi.

\bibitem{UnitarityMethod}
Z.~Bern, L.~J.~Dixon, D.~C.~Dunbar and D.~A.~Kosower,
Nucl.\ Phys.\ B {\bf 425}, 217 (1994)\ifpreprint{} [hep-ph/9403226]\fi;
%
Nucl.\ Phys.\ B {\bf 435}, 59 (1995)\ifpreprint{} [hep-ph/9409265]\fi;
%
Phys.\ Lett.\  B {\bf 394}, 105 (1997)\ifpreprint{} [hep-th/9611127]\fi;
%
Z.~Bern and A.~G.~Morgan,
Nucl.\ Phys.\  B {\bf 467}, 479 (1996)
\ifpreprint{} [hep-ph/9511336]\fi;
%
Z.~Bern, L.~J.~Dixon and D.~A.~Kosower,
Nucl.\ Phys.\  B {\bf 513}, 3 (1998)\ifpreprint{} [hep-ph/9708239]\fi;
%
R.~Britto, F.~Cachazo and B.~Feng,
Nucl.\ Phys.\  B {\bf 725}, 275 (2005)\ifpreprint{} [hep-th/0412103]\fi;
%
C.~Anastasiou {\it et al.},
Phys.\ Lett.\  B {\bf 645}, 213 (2007)
\ifpreprint{} [hep-ph/0609191]\fi;
%
R.~Britto and B.~Feng,
JHEP {\bf 0802}, 095 (2008)
\ifpreprint{} [0711.4284 [hep-ph]]\fi.

\bibitem{BCFW}
R.~Britto, F.~Cachazo, B.~Feng and E.~Witten,
Phys.\ Rev.\ Lett.\ {\bf 94}, 181602 (2005)\ifpreprint{} [hep-th/0501052]\fi.

\bibitem{Bootstrap}
C.~F.~Berger {\it et al.},
Phys.\ Rev.\ D {\bf 74}, 036009 (2006)\ifpreprint{} [hep-ph/0604195]\fi.

\bibitem{OPPEtc}
G.~Ossola, C.~G.~Papadopoulos and R.~Pittau,
Nucl.\ Phys.\  B {\bf 763}, 147 (2007)\ifpreprint{} [hep-ph/0609007]\fi;
%
D.~Forde,
Phys.\ Rev.\  D {\bf 75}, 125019 (2007)\ifpreprint{} [0704.1835 [hep-ph]]\fi;
%
W.~T.~Giele, Z.~Kunszt and K.~Melnikov,
JHEP {\bf 0804} (2008) 049
\ifpreprint{} [0801.2237 [hep-ph]]\fi;
%
S.~D.~Badger,
JHEP {\bf 0901}, 049 (2009)\ifpreprint{} [0806.4600 [hep-ph]]\fi.

\bibitem{D0W123}
V.~M.~Abazov {\it et al.} [D0 Collaboration],
1106.1457 [hep-ex].

\bibitem{TevatronPapers}
T.~Aaltonen {\em et al.} [CDF Collaboration],
Phys.\ Rev.\ Lett.\  {\bf 100}, 102001 (2008)\ifpreprint{} [0711.3717 [hep-ex]]\fi;
%
Phys.\ Rev.\  D {\bf 77}, 011108 (2008)\ifpreprint{} [0711.4044 [hep-ex]]\fi;
%
V.~M.~Abazov {\it et al.}  [D0 Collaboration],
Phys.\ Lett.\  B {\bf 678}, 45 (2009)\ifpreprint{} [0903.1748 [hep-ex]]\fi.

\bibitem{BlackHatI}
C.~F.~Berger {\it et al.},
Phys.\ Rev.\ D {\bf 78}, 036003 (2008)\ifpreprint{} [0803.4180 [hep-ph]]\fi.

\bibitem{HaraldReview}
H. Ita, to appear.

\bibitem{CS}
S.~Catani and M.~H.~Seymour,
Nucl.\ Phys.\  B {\bf 485}, 291 (1997)
[Erratum-ibid.\  B {\bf 510}, 503 (1998)]\ifpreprint{} [hep-ph/9605323]\fi.

\bibitem{Amegic}
F.~Krauss, R.~Kuhn and G.~Soff,
JHEP {\bf 0202}, 044 (2002)\ifpreprint{} [hep-ph/0109036]\fi;
%
T.~Gleisberg and F.~Krauss,
Eur.\ Phys.\ J.\  C {\bf 53}, 501 (2008)\ifpreprint{} [0709.2881 [hep-ph]]\fi.

\bibitem{Sherpa}
T.~Gleisberg {\it et al.},
JHEP {\bf 0902}, 007 (2009)\ifpreprint{} [0811.4622 [hep-ph]]\fi.

\bibitem{AntennaIntegrator}
A.~van Hameren and C.~G.~Papadopoulos,
Eur.\ Phys.\ J.\  C {\bf 25}, 563 (2002)\ifpreprint{} [hep-ph/0204055]\fi.

\bibitem{GleisbergIntegrator}
T.~Gleisberg, S.~H\"oche and F.~Krauss,
0808.3672 [hep-ph].

\bibitem{DrummondTrees}
L.~J.~Dixon, J.~M.~Henn, J.~Plefka and T.~Schuster,
JHEP {\bf 1101}, 035 (2011)
\ifpreprint{} [1010.3991 [hep-ph]]\fi.

\bibitem{MCFM}
J.~M.~Campbell and R.~K.~Ellis,
Phys.\ Rev.\  D {\bf 65}, 113007 (2002)
\ifpreprint{} [hep-ph/0202176]\fi.

\bibitem{deFV}
D.~de Florian and W.~Vogelsang,
Phys.\ Rev.\  D {\bf 76}, 074031 (2007)\ifpreprint{} [0704.1677 [hep-ph]]\fi.

\bibitem{antikT}
M.~Cacciari, G.~P.~Salam and G.~Soyez,
JHEP {\bf 0804}, 063 (2008)\ifpreprint{} [0802.1189 [hep-ph]]\fi.

\bibitem{CTEQ6M}
J.~Pumplin {\it et al.},
JHEP {\bf 0207}, 012 (2002)\ifpreprint{} [hep-ph/0201195]\fi.

\bibitem{BerendsRatio}
S.~D.~Ellis, R.~Kleiss and W.~J.~Stirling,
Phys.\ Lett.\  B {\bf 154}, 435 (1985);
%
F.~A.~Berends {\it et al.},
Phys.\ Lett.\  B {\bf 224}, 237 (1989);
%
F.~A.~Berends, H.~Kuijf, B.~Tausk and W.~T.~Giele,
Nucl.\ Phys.\  B {\bf 357}, 32 (1991);
%
E.~Abouzaid and H.~J.~Frisch,
Phys.\ Rev.\  D {\bf 68}, 033014 (2003)\ifpreprint{} [hep-ph/0303088]\fi.

\bibitem{POWHEGBOXMCNLO}
S.~Frixione and B.~R.~Webber,
JHEP {\bf 0206}, 029 (2002)\ifpreprint{} [hep-ph/0204244]\fi;
P.~Nason,
JHEP {\bf 0411}, 040 (2004)\ifpreprint{} [hep-ph/0409146]\fi;
%
S.~Frixione, P.~Nason and C.~Oleari,
JHEP {\bf 0711}, 070 (2007)\ifpreprint{} [0709.2092 [hep-ph]]\fi;
%
S.~Alioli, P.~Nason, C.~Oleari and E.~Re,
JHEP {\bf 1006}, 043 (2010)\ifpreprint{} [1002.2581 [hep-ph]]\fi;
%
K.~Hamilton and P.~Nason,
JHEP {\bf 1006}, 039 (2010)\ifpreprint{} [1004.1764 [hep-ph]]\fi;
%
S.~H\"oche, F.~Krauss, M.~Sch\"onherr and F.~Siegert,
JHEP {\bf 1104}, 024 (2011)\ifpreprint{} [1008.5399 [hep-ph]]\fi;
%
1009.1127 [hep-ph].

\end{thebibliography}
\end{document}